\documentclass{elsart}
\journal{Physics Letter B}
\usepackage{graphics}
\usepackage{color}

\makeatletter
\DeclareRobustCommand*\cal{\@fontswitch\relax\mathcal}
\DeclareRobustCommand*\mit{\@fontswitch\relax\mathnormal}

\def\endequation{\eqno \hbox{\@eqnnum}$$\@ignoretrue}
\def\eqnarray{%
   \stepcounter{equation}%
   \def\@currentlabel{\p@equation\theequation}%
   \global\@eqnswtrue
   \m@th
   \global\@eqcnt\z@
   \tabskip\@centering
   \let\\\@eqncr
   $$\everycr{}\halign to\displaywidth\bgroup
       \hskip\@centering$\displaystyle\tabskip\z@skip{##}$\@eqnsel
      &\global\@eqcnt\@ne\hskip \tw@\arraycolsep \hfil${##}$\hfil
      &\global\@eqcnt\tw@ \hskip \tw@\arraycolsep
         $\displaystyle{##}$\hfil\tabskip\@centering
      &\global\@eqcnt\thr@@ \hb@xt@\z@\bgroup\hss##\egroup
         \tabskip\z@skip
      \cr
}
\def\endeqnarray{%
      \@@eqncr
      \egroup
      \global\advance\c@equation\m@ne
   $$\@ignoretrue
}
\makeatother

\begin{document}
\begin{frontmatter}
\begin{flushleft}
\hspace*{9cm}BELLE Preprint 2008-28 \\
\hspace*{9cm}KEK Preprint 2008-38 \\
\hspace*{9cm}NTLP 2008-03 \\
\hspace*{9cm}arXiv:0811.0088 [hep-ex] \\
\end{flushleft}

\title{Precise measurement of hadronic $\tau$-decays with an $\eta$ meson}

\collab{Belle Collaboration}
   \author[Nagoya]{K.~Inami}, 
   \author[Nagoya]{T.~Ohshima}, 
   \author[Nagoya]{H.~Kaji}, 
   \author[KEK]{I.~Adachi}, 
   \author[Tokyo]{H.~Aihara}, 
   \author[BINP]{K.~Arinstein}, 
   \author[BINP]{V.~Aulchenko}, 
   \author[Lausanne,ITEP]{T.~Aushev}, 
   \author[BINP]{I.~Bedny}, 
   \author[Panjab]{V.~Bhardwaj}, 
% ...

\address[BINP]{Budker Institute of Nuclear Physics, Novosibirsk, Russia}
% ...
\address[KEK]{High Energy Accelerator Research Organization (KEK), Tsukuba, Japan}
% ...
\address[ITEP]{Institute for Theoretical and Experimental Physics, Moscow, Russia}
% ...
\address[Lausanne]{\'Ecole Polytechnique F\'ed\'erale de Lausanne, EPFL, Lausanne, Switzerland}
% ...
\address[Nagoya]{Nagoya University, Nagoya, Japan}
% ...
\address[Panjab]{Panjab University, Chandigarh, India}
% ...
\address[Tokyo]{Department of Physics, University of Tokyo, Tokyo, Japan}
% ...


\begin{abstract}
We have studied hadronic $\tau$ decay modes involving 
an $\eta$ meson using 490 fb$^{-1}$ of data collected with the Belle 
detector at the KEKB asymmetric-energy $e^+e^-$ collider.
The following branching fractions have been measured:
${\cal B}(\tau^- \to K^- \eta \nu_{\tau})=
(1.58\pm 0.05 \pm 0.09)\times 10^{-4}$, 
${\cal B}(\tau^- \to K^- \pi^0 \eta \nu_{\tau})=
(4.6\pm 1.1 \pm 0.4)\times 10^{-5}$, 
${\cal B}(\tau^- \to \pi^- \pi^0 \eta \nu_{\tau})=
(1.35 \pm 0.03 \pm 0.07) \times 10^{-3}$,
${\cal B}(\tau^- \to \pi^- K_S^0 \eta \nu_{\tau})=
(4.4 \pm 0.7 \pm 0.2) \times 10^{-5}$, and 
${\cal B}(\tau^- \to K^{*-} \eta \nu_{\tau})=
(1.34\pm 0.12 \pm 0.09)\times 10^{-4}$. 
These results are substantially more precise than previous measurements.
The new measurements are compared with theoretical calculations based on the CVC
hypothesis or the chiral perturbation theory. 
We also set upper limits on branching fractions for $\tau$ decays
into $K^- K_S^0 \eta \nu_\tau$, $\pi^- K_S^0 \pi^0 \eta \nu_\tau$,
$K^- \eta \eta \nu_\tau$, $\pi^- \eta \eta \nu_\tau$
and non-resonant $K^- \pi^0 \eta \nu_\tau$ final states.
\end{abstract}

\end{frontmatter}

\section{Introduction}

Hadronic decays of the $\tau$ lepton are very important for studying 
QCD phenomena at a low-energy scale.
Various decay modes including an $\eta$-meson represent a wide class
of such decays, which are still poorly studied. 
One of the important tasks in this field is to test
effective chiral theories \cite{WZ,Witten,Pich,Li}. 
The $\tau$ decay amplitudes in these theories
are based on an effective Lagrangian of pseudoscalar, 
vector and axial-vector mesons with $U(3)_L \times U(3)_R$
symmetry. 
On the other hand, for some decays the branching fractions 
as well as spectral functions of the final states can be predicted 
using the vector dominance model (VDM).
Another important issue is testing the relations
between the cross section of $e^+e^- \to \pi^+ \pi^- \eta$ and 
the spectral function in $\tau^- \to \pi^- \pi^0 \eta \nu_{\tau}$ decay
predicted by conservation of vector current (CVC) and
isospin symmetry~\cite{CVC}.
For all of the above tasks, precise measurements of the branching
fractions as well as detailed studies of the final-state invariant mass
distributions are highly desirable.

Three $\tau$-decay modes involving $\eta$ mesons, 
$\tau^- \to K^- \eta \nu_{\tau}$\footnote{Unless specified otherwise, 
charge-conjugate decays are implied throughout the paper.},
$\tau^- \to K^*(892)^- \eta \nu_{\tau}$, and 
$\tau^- \to \pi^- \pi^0 \eta \nu_{\tau}$, have been studied by 
CLEO~\cite{CLEOpipi0etanu,CLEOketanu,CLEOkpi0etanu} and ALEPH~\cite{ALEPH}. 
However, the experimental statistics were very limited.
For example, CLEO has the most precise measurements with 85
events in the $\tau^- \to K^- \eta \nu_{\tau}$ mode, 125 events in the 
$\tau^- \to \pi^- \pi^0 \eta \nu_{\tau}$ mode and 25 in the 
$\tau^- \to K^{*-} \eta \nu_{\tau}$ mode.
The limited statistics of these measurements result in
large uncertainties for the branching fractions and do not
allow study of the hadronic mass spectra.

We report a detailed study of the branching 
fractions for $\tau^- \to K^- \eta \nu_{\tau}$, 
$K^-\pi^0 \eta \nu_{\tau}$, $\pi^- \pi^0 \eta \nu_{\tau}$, 
as well as $\pi^- K_S^0 \eta \nu_{\tau}$ decays. The $K^-\pi^0$ invariant mass 
in the $K^- \pi^0 \eta \nu_{\tau}$ decay and the $\pi^- K_S^0$ invariant 
mass in the $\pi^- K_S^0 \eta \nu_{\tau}$ decay are studied to evaluate 
the branching fraction of the $\tau^-\to K^*(892)^- \eta \nu_{\tau}$ decay.
In addition, some other final-state invariant mass distributions are 
presented and discussed.

This study was performed at the KEKB asymmetric-energy $e^+e^-$ 
collider~\cite{KEKB} with the Belle detector. 
The results are based on a 490 fb$^{-1}$ data sample that
contains about 450 million $\tau$-pairs, which is almost 100 times larger 
than in any of the previous measurements.

The structure of this paper is as follows.
Section~\ref{BelleDet} describes the Belle detector. 
Event selection is explained in Section~\ref{EventSelection}. 
The branching fractions are determined in Section~\ref{Br1}, 
where that of $\tau^-\to K^*(892)^- \eta \nu_{\tau}$ is evaluated from 
analysis of the $\tau^-\to K^-\pi^0 \eta \nu_{\tau}$ and 
$\pi^- K_S^0 \eta \nu_{\tau}$ samples in Section~\ref{Br2}. Finally, 
some concluding remarks are given in Section~\ref{conclusion}. 

\section{\bf Experimental apparatus and Monte-Carlo simulation}
\label{BelleDet}

The Belle detector is a large-solid-angle magnetic spectrometer that
consists of a silicon vertex detector (SVD),
a 50-layer central drift chamber (CDC), an array of
aerogel threshold Cherenkov counters (ACC),
a barrel-like arrangement of time-of-flight
scintillation counters (TOF), and an electromagnetic calorimeter (ECL)
comprised of CsI(Tl) crystals located inside
a superconducting solenoid coil that provides a 1.5~T
magnetic field.  An iron flux-return located outside 
the coil is instrumented to detect $K_L^0$ mesons and to identify
muons (KLM).  The detector is described in detail elsewhere~\cite{BelleDet}.
Two inner detector configurations were used. A 2.0 cm radius beam pipe
and a 3-layer silicon vertex detector were used for the first sample
of 140~fb$^{-1}$, while a 1.5 cm radius beam pipe, a 4-layer
silicon detector and a small-cell inner drift chamber were used to record
the remaining 350~fb$^{-1}$~\cite{svd2}.

Particle identification (PID) of the charged tracks uses 
likelihood ratios, ${\cal P}_x$, for a charged particle of species $x$
($x=\mu,~e,~K,~\pi~\textrm{or}~p$).
${\cal P}_x$ is defined as ${\cal P}_x=L_x /\Sigma_y L_y$
(the sum runs over the relevant particle species), where $L_x$ is 
a likelihood based on the energy deposit and shower shape in the ECL, 
the momentum and $dE/dX$ measured in the CDC, the particle range 
in the KLM, the light yield in the ACC, and the particle's time-of-flight 
from the TOF counter~\cite{EMu}.
The efficiencies for kaon, muon and electron identification are
77\% for momenta of $p>0.3$ GeV/$c$,
92\% for $p>1.0$ GeV/$c$ and
94\% for $p>0.5$ GeV/$c$, respectively.

The detection efficiency for each signal mode and the level of background (BG)
contribution are estimated from Monte Carlo (MC) 
simulations. To generate signal events as well as $\tau$-decay 
backgrounds, the KKMC program \cite{KKMC} is used. 
For the decay modes not covered by KKMC, like
$\tau^- \to K^- \eta \nu_\tau$, $K^*(892)^- \eta \nu_\tau$, $\ldots$,
the signal MC is produced assuming a pure phase-space distribution
of the hadronic system with a $V-A$ weak interaction.
The background 
from $e^+e^-\to q\bar{q}$ ($q\bar{q}$ continuum) is simulated 
using the EvtGen procedure \cite{EvtGen}.
The detector response is simulated by a GEANT3 \cite{GEANT} based program.
Throughout the paper the efficiencies quoted include the branching ratios
of the corresponding $\eta$ decays.

\section{Event Selections}\label{EventSelection}

The signal events should have the following common features 
in the $e^+ e^- \to \tau^-_{\rm signal} \tau^+_{\rm tag}$ reaction: 
\begin{eqnarray}
\hspace*{0.5 cm} \tau^-_{\rm signal} &\to& X^- + \eta + n(\leq 1)\gamma + 
{\rm (missing)}, \nonumber \\ 
\tau^+_{\rm tag} &\to& (\mu/ e)^+ +  n(\leq 1)\gamma + {\rm (missing)}, 
\nonumber
\end{eqnarray}
where $X^-$ denotes 
$K^-$, $K^-\pi^0$, $\pi^-\pi^0$, $\pi^-K_S^0$ or
$K^*(892)^-$ systems. The $\pi^0$, $K_S^0$ and $K^*(892)^-$ are 
reconstructed through $\pi^0\to \gamma\gamma$, $K_S^0 \to \pi^+\pi^-$, 
and $K^*(892)^-\to K_S^0 \pi^-, K^-\pi^0$ decays, respectively. 
The $\eta$ meson is identified through the 
$\eta\to \gamma\gamma$ decay mode for all $\tau$ channels while
the $\eta\to \pi^+\pi^-\pi^0$ mode is added for the 
$\tau^-\to K^-\eta\nu$ decay mode.  

The signal events, therefore, comprise either two or four charged tracks 
with zero net charge and two or four $\gamma$'s.
In order to take into account initial-state radiation, 
we allow at most one 
extra $\gamma$ on both the signal and tag sides. 
Each charged track is required to have a transverse momentum  
$p_t> 0.1$ GeV/$c$, and a polar angle of $-0.866 < \cos\theta < 0.956$, 
where $p_t$ and $\theta$ are measured relative to the direction opposite 
to that of the incident $e^+$ beam in the laboratory frame. 
Each photon candidate should have an energy $E_{\gamma} > 0.05$ GeV within the 
same polar angle region as for the charged particles.

The tag-side $\tau_{\rm tag}$ is required to decay 
into leptons, i.e. 
$\tau^+\to\ell^+\nu_{\ell}\bar{\nu}_{\tau}$ $(\ell=e/\mu)$, 
which corresponds to a branching fraction of 35.2\%~\cite{PDG}. 
The lepton candidates are required to be well identified 
by requirements on the ${\cal {P}}_{e}$ or ${\cal {P}}_{\mu}$ PID parameters
with $p > 0.7$ GeV/$c$.

The thrust axis ($\textbf{n}_{\rm thrust}$) is defined to maximize 
$V_{\rm thrust}=\Sigma_i |\textbf{p}_i^{\rm CM} 
\cdot \textbf{n}_{\rm thrust}|/\Sigma_i |\textbf{p}_i^{\rm CM}|$, 
where $\textbf{p}_i^{\rm CM}$ denotes the momentum of each particle in 
the center-of-mass system (CM),
and the sum runs over all detected particles in an event. 
The requirement $V_{\rm thrust}>0.8$ is applied to remove $q\bar{q}$ events.
An event is divided into two hemispheres by the plane perpendicular to 
$\textbf{n}_{\rm thrust}$.
The hemisphere
that includes $\tau_{\rm signal}$ with an $\eta$ 
is referred to as the signal side, while the opposite hemisphere, 
which includes $\tau_{\rm tag}$, is defined as the tag side. 

In order to remove non-$\tau$ pair backgrounds, such as the $q\overline{q}$
continuum, the total energy in CM is required to be in the range
$3.0~\textrm{GeV}< E^{\rm CM}_{\rm total} < 10.0$ GeV. 
In addition, the invariant masses of both the signal and the tag sides are
required to be 
smaller than the $\tau$-mass: $M_{\rm sig},~M_{\rm tag} < m_{\tau}$ 
(1.78 GeV/$c^2$).  
To remove two-photon backgrounds, the missing momentum should 
correspond to a particle crossing the fiducial region;
we require that the missing momentum satisfy
$-0.866 < \cos\theta(p_{\rm miss}) < 0.956$. 
After applying these requirements, the $q\bar{q}$ background level
is 3\% of the signal yield.
Background from two-photon reactions is negligible.

To reconstruct an $\eta$, two $\gamma$'s with $E_{\gamma} > 0.2$ GeV 
are required in the barrel region ($-0.624 < \cos\theta < 0.833$). 
In addition, we allow combinations with 
at most one extra $\gamma$ with $0.05 < E_{\gamma} < 0.2$ GeV in 
the endcap region on the signal side
to take into account initial-state 
radiation as well as beam-induced BG clusters in the calorimeter. 
In order to reduce the number of incorrect combinations with 
a $\gamma$ from $\pi^0$ 
decay (denoted hereafter as $\gamma_{\pi^0}$), the $\eta$-candidate 
$\gamma$ ($\gamma_{\eta}$) should not form a $\pi^0$ mass with any 
other $\gamma$, i.e., a $\pi^0$-veto is applied.
The $\pi^0$ mass window is defined in this paper as 
0.105 GeV/$c^2 < M_{\gamma\gamma} <$ 0.165 GeV/$c^2$, which is a 
$\pm 3 \sigma$ range of the detector resolution.

\subsection{$\tau^-\to K^-\eta\nu_{\tau}$ decay}

Candidate events for this decay mode must contain   
one charged track and at least two $\gamma$'s in the $\eta\to\gamma\gamma$ case,
or three charged tracks and two $\gamma$'s in the $\eta\to\pi^+\pi^-\pi^0$ 
$(\pi^0\to\gamma\gamma)$ case. 

\begin{figure}[t]
\centerline{
\resizebox{0.7\textwidth}{!}{%
\includegraphics{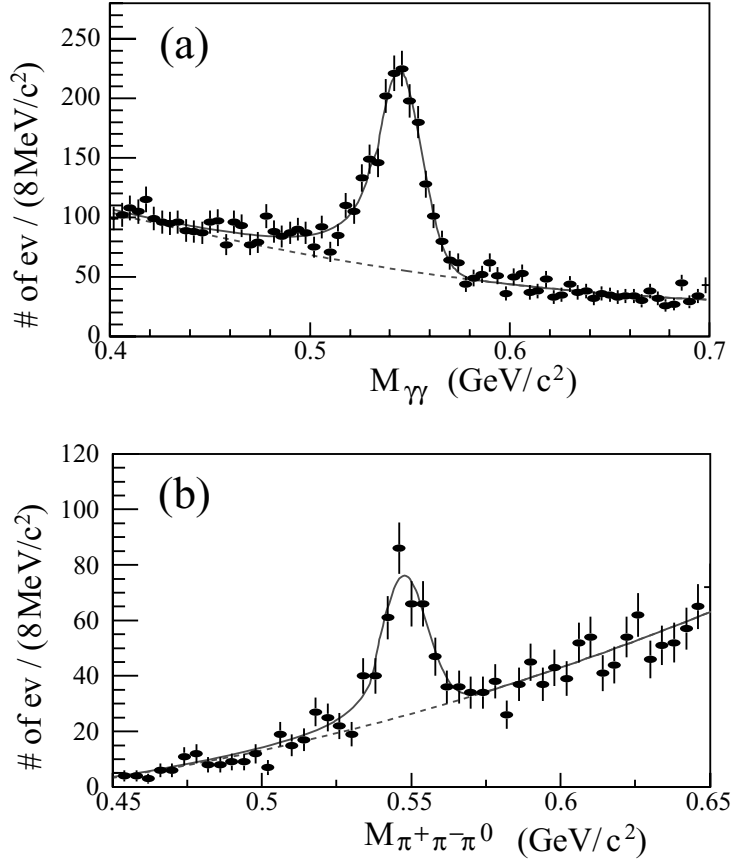}}}
\caption{(a) $M_{\gamma\gamma}$
and (b) $M_{\pi^+\pi^-\pi^0}$ distributions for 
$K^-\eta\nu_{\tau}$ selection in 
$\eta\to\gamma\gamma$ and $\eta\to\pi^-\pi^+\pi^0$ decays,
respectively.
Data are fit with a Crystal Ball function 
plus a second-order polynomial for the BG. 
The best-fit result is indicated by the solid curve with the BG shown 
by the dashed curve.}
\label{fig:fitdata.eps}
\end{figure}

A kaon is identified by the ${\cal {P}}_{K}$ parameter with
$p > 0.3$ GeV/$c$. In addition,
a low ${\cal {P}}_{e}$ value is required to remove beam electrons
from two-photon processes;
this requirement reduces two-photon BG to a negligible level.
The opening angle between the $K^-$ and $\eta$ is required to satisfy 
the condition $\cos\theta(P^{\rm CM}_{K}; P^{\rm CM}_{\eta}) > 0.8$,
to reduce the combinatorial BG. 
In addition, the opening angle and energy of the two $\gamma_{\eta}$'s should 
fulfill the following conditions:
$0.5 < \cos\theta(P^{\rm CM}_{\gamma 1}; P^{\rm CM}_{\gamma 2}) < 0.96$ and  
$|E^{\rm CM}_{\gamma 1}-E^{\rm CM}_{\gamma 2}|/|E^{\rm CM}_{\gamma 1}+
E^{\rm CM}_{\gamma 2}| < 0.8$, respectively.

For $\eta\to\pi^+\pi^-\pi^0$ reconstruction, 
the selection criteria are basically the same as in 
the $\eta\to \gamma\gamma$ case. 
Here $\eta$ candidates should have two additional charged tracks, 
and two of the $\gamma$'s must form a $\pi^0$ instead of an $\eta$. 

The $\gamma \gamma$ and $\pi^+ \pi^- \pi^0$ invariant mass distributions
around $\eta$ mass 
are shown in Figs.\ref{fig:fitdata.eps} (a) and (b), respectively.
The $\eta$ peak is clearly seen in both cases. 
The selection efficiencies (including the intermediate branching fractions 
of ${\cal B}(\eta\to\gamma\gamma)$, ${\cal B}(\eta\to\pi^+\pi^-\pi^0)$, and
${\cal B}(\tau^-\to \ell^-\nu_{\tau}\bar{\nu}_{\ell})$)
are $\epsilon=0.94\%$ in the $\eta\to\gamma\gamma$ and 
$\epsilon=0.16\%$ in the $\eta\to\pi^+\pi^-\pi^0$ case.

\subsection{$\tau^-\to K^- \pi^0 \eta \nu_{\tau}$ and 
$\pi^-\pi^0\eta\nu_{\tau}$ decays} 

For these decays, an $\eta$ is reconstructed through the $\gamma\gamma$ mode
only, so that the signal event should contain one charged track and four 
$\gamma$'s on the signal side. 

All $\gamma$'s, except for a possible extra $\gamma$, are required to 
be detected in
the barrel region of the calorimeter, and two of them should
form a combination consistent the $\pi^0$ mass.  
The total momentum on the signal side is required to satisfy 
$\sum p^{\rm CM}_{\rm sig} > 2.5$ GeV/$c$. 
A condition on the opening angle between the missing momentum 
in the CM and the thrust axis is imposed: 
$\cos\theta(\mathbf{p}^{\rm CM}_{\rm miss}; 
\mathbf{n}^{\rm sig}_{\rm thrust}) < -0.6$,
to remove BG's with large missing
momentum on the signal side. 
Charged kaons and pions are selected by requiring 
large ${\cal P}_{K}$ and small ${\cal P}_e$
in the $K^-\pi^0\eta\nu_{\tau}$ mode, while 
small ${\cal P}_{K}$ and small ${\cal P}_e$ are selected in 
the $\pi^-\pi^0\eta\nu_{\tau}$ mode. 

\begin{figure}[t]
\centerline{
\resizebox{0.9\textwidth}{!}{%
\includegraphics{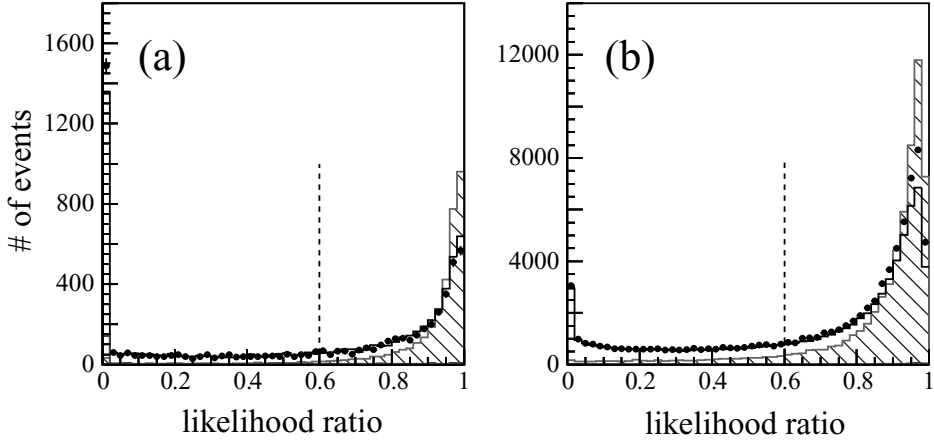}}}
  \caption{Distribution of the likelihood ratio for 
  (a) $K^-\pi^0\eta\nu_{\tau}$ and (b) $\pi^-\pi^0\eta\nu_{\tau}$. 
  Dots are experimental data. 
  The hatched and normal histograms 
  indicate the signal and BG
  MC distributions, respectively. The dashed line shows the threshold of 
  this selection.}
  \label{fig:like.eps}
\end{figure}

For further background rejection, we construct a likelihood
using the following seven variables: 
$V_{\rm thrust}$, momenta of the $\eta$ and $\pi^0$ in the CM 
($p^{\rm CM}_{\eta}$, $p^{\rm CM}_{\pi^0}$), the missing-mass squared 
($M^2_{\rm miss}$), the energy of the $\gamma_\eta$'s in the CM 
($E^{\rm CM}_{\gamma_{\eta}}$), $\sum p^{\rm CM}_{\rm sig}$, and 
$\cos\theta(p^{\rm CM}_{K/\pi}; p^{\rm CM}_{\eta})$. 
The MC simulation is used to study the likelihood distributions of 
generic $\tau^+\tau^-$ and the small residual
$q\bar{q}$ BG (${\it L}_{\rm BG}$) 
as well as the signal (${\it L}_{\rm sig}$). 
A likelihood ratio is defined as 
${\it R}={\it L}_{\rm sig}/({\it L}_{\rm sig}+{\it L}_{\rm BG})$.  
The ${\it R}$ distribution is shown in Figs.~\ref{fig:like.eps} (a) for 
$K^-\pi^0\eta\nu_{\tau}$  and (b) for $\pi^-\pi^0\eta\nu_{\tau}$ decay; 
with the requirement ${\it R} > 0.6$, about half of the background is removed,
while 93\% and 90\% of the signal samples are retained, respectively. 

\begin{figure}[t]
\centerline{
\resizebox{0.7\textwidth}{!}{%
\includegraphics{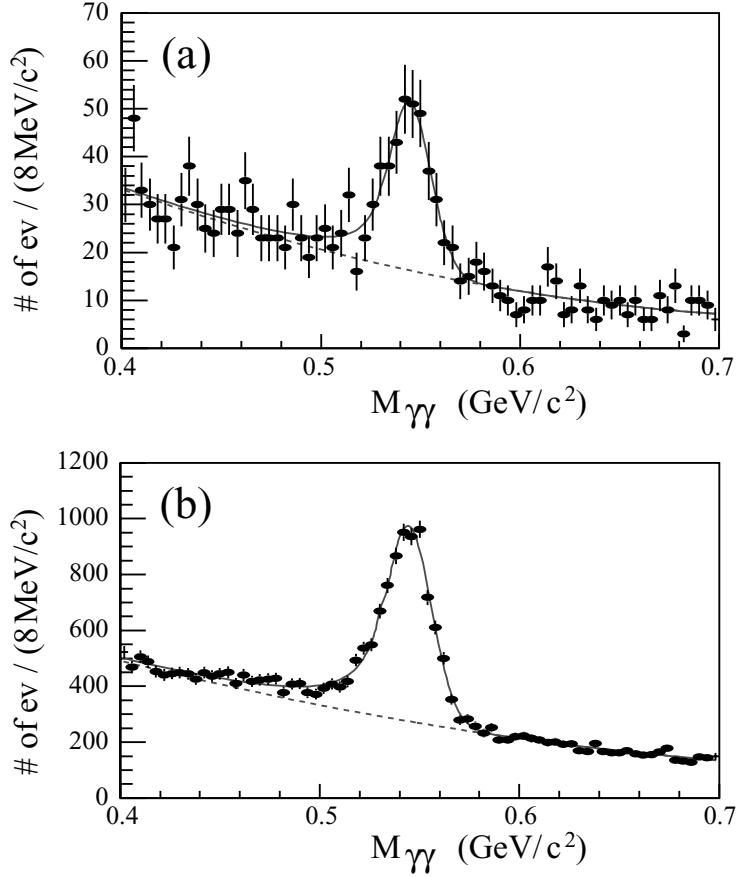}}}
\caption{$M_{\gamma\gamma}$ distributions 
for (a) $K^-\pi^0\eta\nu_{\tau}$ and (b) $\pi^-\pi^0\eta\nu_{\tau}$ 
candidates. 
Data are fit with a Crystal Ball function plus a second-order polynomial 
for the BG.
The result of the best fit is indicated by the solid curve with the BG 
shown by the dashed curve.}
\label{fig:fig2.eps}
\end{figure}

The resulting $M_{\gamma\gamma}$ distributions are shown in 
Figs.\ref{fig:fig2.eps} (a) and (b) for the $K^-\pi^0\eta\nu_{\tau}$ 
and $\pi^-\pi^0\eta\nu_{\tau}$ modes, respectively.  
The $\eta$ peaks are clearly visible.
The detection efficiencies are $\epsilon = 0.35 \%$ and $0.47 \%$ 
in the $\tau^-\to K^-\pi^0\eta\nu_{\tau}$ and $\pi^-\pi^0\eta\nu_{\tau}$ 
decays, respectively.

\subsection{$\pi^- K_S^0 \eta\nu_{\tau}$ decay}

On the signal side, three charged tracks and at least two $\gamma$'s are 
required.
To suppress two-photon BG, the tracks should not be identified
as electrons.
Without $K/\pi$ ID the pion mass of $m_{\pi^{\pm}}=0.13957$ GeV/$c^2$ is 
assigned to all three tracks on the signal side, two of the three tracks 
with opposite charges are the $K_S^0$ daughters, 
the invariant mass ($M_{\pi^+\pi^-}$) of which 
is required to be within a $K_S^0$ mass window ($3\sigma$), 
$0.4907 \ \textrm{GeV/}c^2<M_{\pi^+\pi^-}<0.5047 \ \textrm{GeV/}c^2$. 
To ensure that these tracks are $K_S^0$ daughters,
the distance from the interaction 
point to the $K_S^0$ decay vertex on the plane perpendicular to the beam axis 
($r_{\rm vertex}$) is required to satisfy the condition: 
$0.5 \ \textrm{cm}<r_{\rm vertex}<30 \ \textrm{cm}$. 
To avoid multiple candidates, the mass of the other possible combination must 
not satisfy the $K_S^0$ mass condition. 
Pairs of $\gamma$'s with $E_{\gamma}>0.3$ GeV are used to form $\eta$
candidates. We require that
the energy of an $\eta$ candidate, $E_{\eta}$, must be greater than 1.2 GeV.

The resulting $M_{\gamma\gamma}$ distribution is shown in 
Fig.\ref{fig:fig3.eps} (a). 
The detection efficiency is $\epsilon=0.32 \%$,
which includes the tag-side efficiency,
${\cal B}(\eta \to \gamma \gamma)$ and ${\cal B}(K_S^0 \to \pi^+ \pi^-)$.

\section{Branching Fractions}\label{Br1}

\subsection{$K^-\eta\nu_{\tau}$, $K^-\pi^0\eta\nu_{\tau}$ 
and $\pi^-\pi^0\eta\nu_{\tau}$ decays}

\subsubsection{Signal candidates}\label{subsub311}

The $\eta$ meson yields are obtained from fits to 
the $\gamma \gamma$ invariant mass distributions
shown in Figs.\ref{fig:fitdata.eps}
and \ref{fig:fig2.eps}. 
Fits are performed with a Crystal Ball (CB) function~\cite{CB}
plus a second-order polynomial function to represent the $\eta$ 
contribution and combinatorial BG, respectively. 
All five parameters of the CB function, as well as three others for BG are 
treated as free parameters in the fit. 
 
The results of the fit are shown by the solid curve in 
the corresponding figures. 
The best fits give $\eta$ masses of 
$M_{\eta} = 0.5449\pm 0.0006$, $0.5444\pm 0.0010$ and 
$0.5443\pm 0.0004$ GeV/$c^2$ for $K^-\eta\nu_{\tau}$, 
$K^-\pi^0\eta\nu_{\tau}$, and $\pi^-\pi^0\eta\nu_{\tau}$ decays, 
respectively, in the $\eta\to\gamma\gamma$ subsample,
while $M_{\eta} = 0.5474\pm 0.0007$ GeV/$c^2$ for $K^-\eta\nu_{\tau}$ 
in the $\eta\to\pi^+\pi^-\pi^0$ subsample.
The corresponding mass resolutions are
$\sigma_{M_{\eta}} = 0.0110\pm 0.0006$, $0.0107\pm 0.0015$, 
$0.0116\pm 0.0003$, and $0.0075\pm 0.0004$ GeV/$c^2$, respectively. 
These values of $M_{\eta}$ and $\sigma_{M_{\eta}}$ are 
in good agreement with those obtained from the MC simulation,
although the obtained masses are shifted by 2 - 3 $\sigma$ from
the $\eta$ mass in Ref.~\cite{PDG}, which is due to the incomplete
detector calibration. 
For instance, $M_{\eta}^{\rm MC} =$ 0.5442, 0.5450, 0.5464, and 0.5473 
GeV/$c^2$ for a generated mass of $m_{\eta}=$ 0.54745 
GeV/$c^2$. 

The $\eta$ yields obtained from the fits are $N_{\eta} = 1,387\pm 43$, 
$270\pm 33$, and $5,959\pm 105$ events for the $K^-\eta\nu_{\tau}$, 
$K^-\pi^0\eta\nu_{\tau}$, and $\pi^-\pi^0\eta\nu_{\tau}$ modes, 
respectively, in the $\eta\to\gamma\gamma$ case, 
while $N_{\eta} = 241\pm 21$ events for the $K^-\eta\nu_{\tau}$ 
decay in the $\eta\to\pi^+\pi^-\pi^0$ case. 
These yields include BG, 
attributed to other $\tau$ decays with $\eta$ meson(s), such as 
the $\tau^- \to \pi^-\pi^0\pi^0\eta\nu_{\tau}$ and 
$\pi^-\pi^+\pi^-\eta\nu_{\tau}$; 
$q\bar{q}$ continuum with $\eta$ meson(s); cross-feeds among the signal 
modes.

\subsubsection{Background}

The possible backgrounds in the $K^- \eta \nu_\tau$, $K^-\pi^0 \eta \nu_\tau$
and $\pi^- \pi^0 \eta \nu_\tau$ decay modes include 
(a) feed-across from signal modes and (b) other processes such as 
$\tau \to \pi^-\pi^0\pi^0\eta\nu_\tau$, $K^-\eta\eta\nu_\tau$, 
$\pi^-\eta\eta\nu_\tau$, $\pi^-\eta\nu_\tau$ and $q\bar{q}$.
We first discuss the second source.

BG from generic $\tau^+\tau^-$ decays is evaluated using a MC simulation
where the value of ${\cal B}(\tau^- \to \pi^-\pi^0\pi^0\eta\nu_\tau)$
is taken from Ref.~\cite{CLEO6pion}.
This BG is negligible for $K^-\eta\nu_{\tau}$ and $K^-\pi^0\eta\nu_{\tau}$ 
decays, and is only $72\pm 20$ events, which is smaller than the statistical 
uncertainty in the $\eta$ yield
for the $\pi^-\pi^0\eta\nu_{\tau}$ decay. 

\begin{figure}[t]
\centerline{
\resizebox{0.9\textwidth}{!}{%
\includegraphics{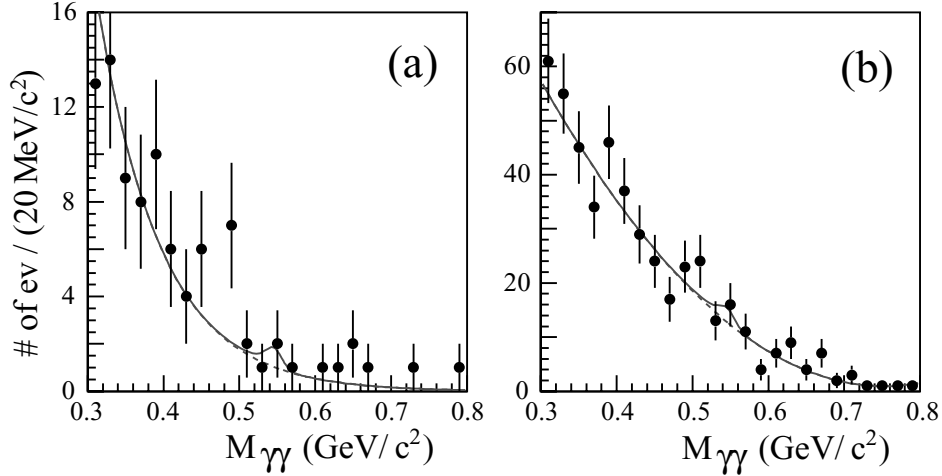}}}
  \caption{Mass of the $\gamma_{\eta}$'s combination,
  which does not include 
  the largest energy $\gamma_{\eta}$, $M_{\gamma\gamma}^{(2)}$ for
  (a) the $\tau^-\to K^-\eta\eta\nu_{\tau}$ sample and 
  (b) the $\tau^-\to \pi^-\eta\eta\nu_{\tau}$ sample. Data are fit with 
  a Crystal Ball function plus a second-order polynomial 
for the BG. The result of the best fit is indicated by 
the solid curve with the BG 
shown by the dashed curve.}
  \label{fig:etaeta.eps}
\end{figure}

To examine the $\tau^-\to K^-\eta\eta\nu_{\tau}$ and 
$\pi^-\eta\eta\nu_{\tau}$ 
contaminations, we analyze data using selection criteria similar to 
those used for the 
$K^-\pi^0\eta\nu_{\tau}$ and $\pi^-\pi^0\eta\nu_{\tau}$ modes.
An additional $\eta$, instead of a $\pi^0$, is reconstructed, 
and no selection on the likelihood ratio is implemented. 
Requiring one photon pair to be in the $\eta$ mass region,
$0.48 {\rm GeV/}c^2 <M_{\gamma \gamma}^{(1)}<0.58$ GeV/$c^2$, 
which includes the largest energy $\gamma$,
the invariant mass distribution of the other $\eta$ candidate,
$(M_{\gamma\gamma}^{(2)})$, is shown in
Figure~\ref{fig:etaeta.eps}.
The evaluated $\eta$ yields are $1.4^{+2.6}_{-1.9}$ and $4.4 \pm 6.8$ events 
in the $K^-\eta\eta\nu_{\tau}$ and $\pi^-\eta\eta\nu_{\tau}$ decays, 
respectively, 
with detection efficiencies of $\epsilon = 0.19\%$ and $0.22\%$,
respectively.
We then set the upper limits~\cite{UL} 
\begin{eqnarray}
{\cal B}(\tau^-\to K^-\eta\eta\nu_{\tau}) < 3.0\times 10^{-6}
\end{eqnarray}
and 
\begin{eqnarray}
{\cal B}(\tau^-\to \pi^-\eta\eta\nu_{\tau}) < 7.4\times 10^{-6} 
\end{eqnarray}
at the 90$\%$ confidence level (CL)
including systematic uncertainties of 6.9\% and 7.1\%, respectively.
Since the cross feed probability for these modes with extra $\eta$'s is
below 3\%, their contamination is ignored. 

The $\tau^-\to \pi^-\eta\nu_{\tau}$ decay proceeds via a second-class 
current so that its branching fraction is expected to be small,
as low as $10^{-5}$~\cite{Pich} or less.
Since the efficiency for detecting $\pi^- \eta \nu_{\tau}$ as signal is 
of the order of $10^{-4}$, the total contamination is then 
$\sim 10^{-9}$ or less so that this decay can be ignored. 

BG from the $q\bar{q}$ continuum is examined using both MC and data 
in order to take into account the $\eta$ production uncertainty of 
the $q\bar{q}$ MC. 
A $q\bar{q}$ enriched sample is produced with some variations of the signal 
selection criteria.
We require 
$M_{\rm tag}>m_{\tau}$, while the requirement $M_{\rm sig} < m_{\tau}$ is 
not applied; the PID requirement on the tag side is reversed.
Other criteria are not changed.
A comparison between the resulting $\eta$ yields of MC and 
the data resulted in the following MC scale factors:
$2.3 \pm 0.7$ for $K^-\eta\nu_{\tau}$, $2.6 \pm 0.5$ for 
$K^-\pi^0\eta\nu_{\tau}$, 
and $2.9 \pm 0.3$ for $\pi^-\pi^0\eta\nu_{\tau}$.
Therefore, we obtain rescaled $q\bar{q}$ contaminations of
$39.7 \pm 15.8$ events in $K^-\eta\nu_{\tau}$, 
$212 \pm 29$ events in $\pi^-\pi^0\eta\nu_{\tau}$ and 
$27.0 \pm 8.5$ events in the $K^-\pi^0\eta\nu_{\tau}$ decay.
These BG's correspond to 2.8\%, 10\% and 3.6\% of the raw signal yields,
$N_{\eta}$, respectively. 

The yields in the three candidate samples
after the BG subtraction evaluated above, $N_i$, 
($i=$ $K^- \eta \nu$, $K^- \pi^0 \eta \nu$, $\pi^- \pi^0 \eta \nu$) are used to
evaluate the corresponding branching fractions, ${\cal B}_i$.
To take into account the cross-feeds between
decay channels, we solve the following system of linear equations:
\begin{eqnarray}\label{eq1}
N_j = 2 N_{\tau\tau} \sum_{i=1}^3 \epsilon_j^i 
{\cal B}_i \hspace*{1.6 cm}(j=1,2,3),
\end{eqnarray}
where 
$\epsilon_j^i$ is the efficiency for detecting mode $j$
as mode $i$, which is estimated by MC;
$N_{\tau\tau}$ is the total number of $\tau$-pairs produced.  
In the $\eta\to\pi^+\pi^-\pi^0$ case, 
${\cal B}(\tau^-\to K^-\eta\nu_{\tau})$ is calculated using 
the relation $N'_1 = 2 N_{\tau\tau} \sum_{i=1}^3 \epsilon'{_1^i} {\cal B}_i$  
(the prime indicates quantities for this $\eta$ mode) 
with ${\cal B}_2$ and ${\cal B}_3$ obtained in the $\eta\to \gamma\gamma$ case. 
The statistical uncertainty in $N'_1$ is $\sim 10\%$, while 
the cross-feed contamination is $\sim 4\%$, and is thus insignificant.  

The resulting branching fractions are 
${\cal B}(\tau^- \to K^-\eta\nu_{\tau})=(1.57\pm 0.05)\times 10^{-4}$, 
${\cal B}(\tau^- \to K^-\pi^0\eta\nu_{\tau})=(4.6\pm 1.1)\times 10^{-5}$, and 
${\cal B}(\tau^- \to \pi^-\pi^0\eta\nu_{\tau})=(1.35\pm 0.03)\times 10^{-3}$ 
in the $\eta\to\gamma\gamma$ case, and 
${\cal B}(\tau^- \to K^-\eta\nu_{\tau})=(1.60\pm 0.15)\times 10^{-4}$ 
in the $\eta\to\pi^+\pi^-\pi^0$ case.

\subsubsection{Systematic uncertainties}\label{sys-1}

Systematic uncertainties are discussed below, and listed in 
Table \ref{Table:sysKetanu}. 
\begin{table}[tb]
 \caption {Summary of systematic uncertainties in each mode ($\%$). 
 The uncertainty in the number of $\tau^+\tau^-$ events comes from 
 the uncertainties in our luminosity measurement by Bhabha events
 and the cross section of $e^+e^-\to \tau^+\tau^-$~\cite{crosssection}.}
 \label{Table:sysKetanu}
 \begin{center}
 \begin{tabular}{c|cccc}\hline
 Signal modes & $K^-\eta\nu$ & $K^-\pi^0\eta\nu$ & $\pi^-\pi^0\eta\nu$ & $K^-\eta\nu$
 \\ 
\hline\hline 
 Items & \multicolumn{3}{|c|}{$\eta\to \gamma\gamma$} & 
                             $\eta\to 3\pi$ \\ 
\hline 
BG subtraction & & & & \\
$K^-\eta\nu_{\tau}$               & $-$ & 0.6 & $1.8\times 10^{-3}$ & $-$ \\
$K^-\pi^0\eta\nu_{\tau}$          & 0.3 & $-$ & $4.2\times 10^{-2}$ & 0.4 \\
$\pi^-\pi^0\eta\nu_{\tau}$        & $7.5\times 10^{-2}$ & 3.3 & $-$ & 0.1 \\
$\pi^-\pi^0\pi^0\eta\nu_{\tau}$   & $-$ & $-$ & 0.4 & $-$ \\
$q\bar{q}$                        & 1.5 & 6.0 & 0.5 & 1.5 \\
\hline
Detection efficiency & & & & \\
$K/\pi$- /lepton-id       &  3.3/ 2.3 & 2.2/ 2.8 & 1.0/2.6 & 2.8/ 2.6 \\
Tracking                 &  1.3 & 1.3 & 1.3 & 3.3 \\
$\pi^0/ \eta \to \gamma\gamma$   &  $-$/ 2.0 & 2.0/ 2.0 & 2.0/ 2.0 & 2.0/ $-$ \\
$\pi^0$ veto             &  2.8 & 2.8 & 2.8 & $-$ \\
\hline
Stat. error of signal MC    &  0.5 & 1.7 & 0.5 & 1.3 \\
${\cal B}(\eta \to \pi^+\pi^-\pi^0)$ & $-$ & $-$ & $-$ & 1.6 \\
\hline
Luminosity meas. & \multicolumn{4}{|c}{1.4}  \\
$\sigma(e^+e^-\to\tau^+\tau^-)$  & \multicolumn{4}{|c}{0.3} \\
\hline \hline
Total                             &  5.9 & 9.1 & 5.3 & 6.2 \\
 \hline
\end{tabular}
 \end{center}
\end{table}

Subtraction of the BG discussed above provides the systematic
uncertainties includes the statistical errors in the
detection efficiencies for BG processes,
as listed in the Table.
Sizable uncertainties are found only in $K^-\pi^0\eta\nu$ decay 
from $\pi^-\pi^0\eta\nu_{\tau}$ and the $q\overline{q}$ continuum 
and amount to 
3.3\% and 6.0\%, respectively.

The systematic uncertainties for PID, track finding, $\pi^0$ and $\eta$ 
reconstruction, and the $\pi^0$-veto are also estimated. 
The modeling of the PID likelihood is tested by studying 
inclusive $D^{*-}$ samples for $K/\pi$ and two-photon 
$\gamma\gamma\to \ell^+\ell^-$ samples for leptons; the 
Uncertainty in the PID efficiency is 2-3\%, 1\% and $\sim 2.5\%$
for $K^{\pm}$, $\pi^{\pm}$ and $(\mu/e)^{\pm}$, respectively. 
The uncertainty for finding a track is 1.0 $\%$ for hadrons and 0.3\% for
leptons. 
The uncertainty in $\pi^0/\eta\to\gamma\gamma$ reconstruction is 2.0\%, 
obtained from a comparison of $\eta\to\gamma\gamma$ and 
$\eta\to\pi^0\pi^0\pi^0$ data samples. 
The $\pi^0$-veto inefficiency in $\pi^-\pi^0\eta\nu_{\tau}$ samples 
is evaluated by comparing its effect on the data and MC samples 
when the tag-side lepton-PID criterion is reversed from the ordinary 
selection. 

The statistical errors on the MC simulation are 0.5$\%$, 1.7$\%$ and 
0.5$\%$ for $\tau^-\to K^-\eta\nu_{\tau}$, $K^-\pi^0\eta\nu_
{\tau}$ and $\pi^-\pi^0\eta\nu_{\tau}$ in the $\eta\to\gamma\gamma$ 
case, respectively, and 1.3\% for $\tau^-\to K^-\eta\nu_{\tau}$ 
in the $\eta\to\pi^+\pi^-\pi^0$ case. 
The branching fractions used in the MC are cited from 
Ref.~\cite{PDG}, where 
${\cal B}(\eta \to \pi^+\pi^-\pi^0)$ yields a sizable uncertainty of  
1.6\%, while others are negligible. 
The uncertainty of the $\sigma(e^+e^-\to\tau^+\tau^-)$ cross section and 
the integrated luminosity are 0.3\% and 1.4\%, respectively. 

All contributions are summed up in quadrature to
obtain the total 
uncertainties; they amount to 5.9\%, 9.1\% and 5.3\% for 
${\cal B}(\tau^-\to K^-\eta\nu_{\tau})$, 
${\cal B}(\tau^-\to K^-\pi^0\eta\nu_{\tau})$ and 
${\cal B}(\tau^-\to \pi^-\pi^0\eta\nu_{\tau})$, respectively, in the 
$\eta\to \gamma\gamma$ case, 
and 6.2\% for ${\cal B}(\tau^-\to K^-\eta\nu_{\tau})$ in the
$\eta\to\pi^+\pi^-\pi^0$ case. 

Taking into account the systematic errors, we obtain the 
following branching fractions: 
\begin{eqnarray}
{\cal B}(\tau^-\to K^-\eta\nu_{\tau})  \hspace*{4.7 cm} \nonumber \\ 
= (1.57\pm 0.05\pm 0.09)\times 10^{-4} ~~~ {\rm for}~ \eta\to\gamma\gamma, \hspace*{0.8 cm}\\
= (1.60\pm 0.15\pm 0.10)\times 10^{-4} ~~~ {\rm for}~ \eta\to\pi^+\pi^-\pi^0, \\
{\cal B}(\tau^-\to K^-\pi^0\eta\nu_{\tau})  \hspace*{4.5 cm} \nonumber \\ 
= (4.6~\pm 1.1\pm 0.4)\times 10^{-5}, \hspace*{2 cm} \\
{\cal B}(\tau^-\to \pi^-\pi^0\eta\nu_{\tau}) \hspace*{4.8 cm} \nonumber \\ 
= (1.35\pm 0.03\pm 0.07)\times 10^{-3}. \hspace*{2 cm} 
\end{eqnarray}
For $K^-\eta\nu_{\tau}$, the two measurements are combined to obtain
\begin{eqnarray}
{\cal B}(\tau^-\to K^-\eta\nu_{\tau}) \hspace*{3.7 cm} \nonumber \\ 
=  (1.58\pm 0.05 \pm 0.09)\times 10^{-4}.
\end{eqnarray}
\vspace*{3 mm}


\subsection{$\pi^-K_S^0\eta\nu_{\tau}$ decay}

\vspace*{2 mm}

\subsubsection{Signal candidates}

\begin{figure}[t]
\centerline{
\resizebox{0.9\textwidth}{!}{%
\includegraphics{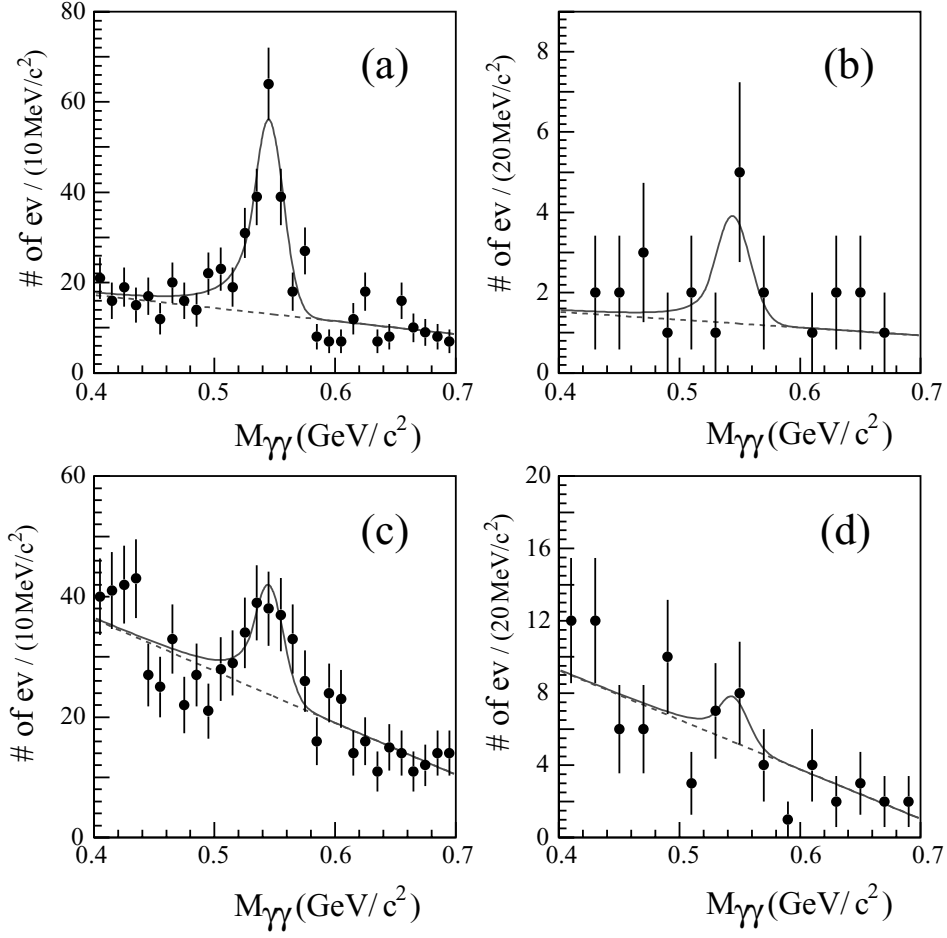}}}
\caption{$M_{\gamma\gamma}$ distributions for 
(a) $\pi^-K_S^0\eta\nu_{\tau}$, (b) $K^-K_S^0\eta\nu_{\tau}$, 
(c) $\pi^-\pi^+\pi^-\eta\nu_{\tau}$ without a $K_S^0$, and 
(d) $\pi^-\pi^0 K_S^0\eta\nu_{\tau}$ samples. 
Data are fit with a Crystal Ball function plus a second-order polynomial 
for the BG. 
The result of the best fit is indicated by the solid curve with the BG 
shown by the dashed curve.}
\label{fig:fig3.eps}
\end{figure}

For $\pi^- K_S^0\eta\nu_{\tau}$, the signal yield $N_{\eta}$ is evaluated from
a fit to the $M_{\gamma\gamma}$ mass distribution in the same way, as discussed
in \ref{subsub311}, but the parameters of the CB function 
are fixed to those determined by the signal MC simulation. 
The fit result is shown in Fig.~\ref{fig:fig3.eps} (a), yielding 
$N_{\eta} = 161 \pm 18$ events. 

We examine $\tau^- \to K^-K_S^0\eta\nu_{\tau}$ and 
$\pi^-\pi^+\pi^-\eta\nu_{\tau}$
decays in order to determine their cross-feed contaminations, with selection 
criteria similar to those for $\pi^-K_S^0\eta\nu_{\tau}$. 
For $K^-K_S^0\eta\nu_{\tau}$, the charged track on the signal side is required 
to be a kaon; $N_{\eta} = 5.6 \pm 3.6$ events is 
obtained from the fit (see Fig.\ref{fig:fig3.eps} (b)).
For $\tau^- \to \pi^-\pi^+\pi^-\eta\nu_{\tau}$, its $M_{\gamma\gamma}$ mass 
distribution, as shown in Fig.~\ref{fig:fig3.eps} (c), is formed from a sample
in which the $\pi^+\pi^-$ invariant mass lies outside the $K_S^0$ mass range
while the other requirements for $K_S^0$ daughters are the same; 
the fit gives $N_{\eta} = 67.9 \pm 17.0$ events.

\subsubsection{Background}

The possible background to the signal comes from
$\pi^-\pi^0 K_S^0\eta\nu_{\tau}$ and the $q\overline{q}$ continuum.
The former sample is selected, requiring an additional $\pi^0$ on 
the signal side. 
No clear $\eta$ peak is found in the $M_{\gamma\gamma}$ distribution, 
as can be seen in Fig.~\ref{fig:fig3.eps} (d):
$N_{\eta} = 4.7^{+6.0}_{-5.1}$ events 
is obtained with a detection efficiency of $\epsilon= 0.08\%$. 
We therefore set an upper limit on the branching fraction
${\cal B}(\tau^-\to \pi^-K_S^0\pi^0\eta\nu_{\tau}) < 2.5\times 10 ^{-5}$ 
at the 90\% CL. 
The BG from this mode in the $\pi^- K_S^0 \eta \nu_{\tau}$ sample is calculated 
to be $1.6^{+2.0}_{-1.7}$ events. 

The contamination of $q\bar{q}$ in
$\pi^- K_S^0 \eta\nu_{\tau}$ decay is $21.1 \pm 7.5$ events, 
following from MC simulation with a scale factor
estimated from the $q\bar{q}$ enriched sample. 

After subtracting the two above BG's from the yield $N_{\eta}$ for
$\pi^- K_S^0 \eta\nu_{\tau}$ decay, we solve the simultaneous equations of 
Eq.(\ref{eq1}) in order to take into account the cross-feeds among 
the three modes. Now, the (super)subscripts $i, j = 1, 2, 3$ for the equations 
correspond to  
$1=\pi^- K_S^0\eta\nu_{\tau}$, $2=K^- K_S^0 \eta\nu_{\tau}$, 
$3=\pi^-\pi^+\pi^-\eta\nu_{\tau}$. 
The last category includes other peaking BG from generic $\tau$ decays, 
which are estimated using samples in the $K_S^0$ sidebands. 
The detection efficiencies are, for instance, 
$\epsilon_1^1=0.32\%$ and $\epsilon_2^2=0.23\%$. 
The resulting branching fractions are 
${\cal B}(\tau^-\to \pi^- K_S^0 \eta\nu_{\tau}) = (4.38\pm 0.75)\times 10^{-5}$
and 
${\cal B}(\tau^-\to K^- K_S^0 \eta\nu_{\tau}) = (1.9\pm 2.0)\times 10^{-6}$,
respectively.

\subsubsection{Systematic uncertainties}

Systematic errors for ${\cal B}(\tau^-\to\pi^- K_S^0\eta\nu_{\tau})$ 
include the following sources.
Uncertainty in the signal CB function used to fit the 
$M_{\gamma\gamma}$ spectrum is examined by varying $M_\eta$ and
$\sigma_{M_\eta}$ within their errors: 
the fitted yield $N_{\eta}$ varies by 0.7\%. 
The dominant BG comes from the $q\bar{q}$ continuum: its uncertainty 
is evaluated to be 6.2\%, while that of the others is negligible.
The uncertainties in the luminosity evaluation and 
$\sigma(e^+e^-\to\tau^+\tau^-)$ are 0.3\% and 1.4\%, respectively. 
The value of 
${\cal B}(K_S^0\to\pi^+\pi^-)$ used in MC has a 0.4\% error~\cite{PDG}. 
The uncertainty in the CB function yields a 0.5\% contribution. 
The total systematic uncertainty is consequently calculated to be
7.9\%, by adding all of the above errors in quadrature.

As a result, we obtain the following branching fractions:
\begin{eqnarray}
{\cal B}(\tau^-\to \pi^- K_S^0 \eta\nu_{\tau}) \hspace*{2.5 cm} \nonumber\\
= (4.4\pm 0.7\pm 0.3)\times 10^{-5}, 
\end{eqnarray}
and
\begin{eqnarray}
{\cal B}(\tau^-\to K^- K_S^0 \eta\nu_{\tau}) &<& 4.5\times 10^{-6},
\end{eqnarray}
at the 90\% CL. 

\vspace*{3 mm}

\subsection{$\tau^- \to K^{*-}\eta\nu_{\tau}$ decay}\label{Br2}
\vspace*{2 mm}

\begin{figure}[t]
\centerline{
\resizebox{0.7\textwidth}{!}{%
\includegraphics{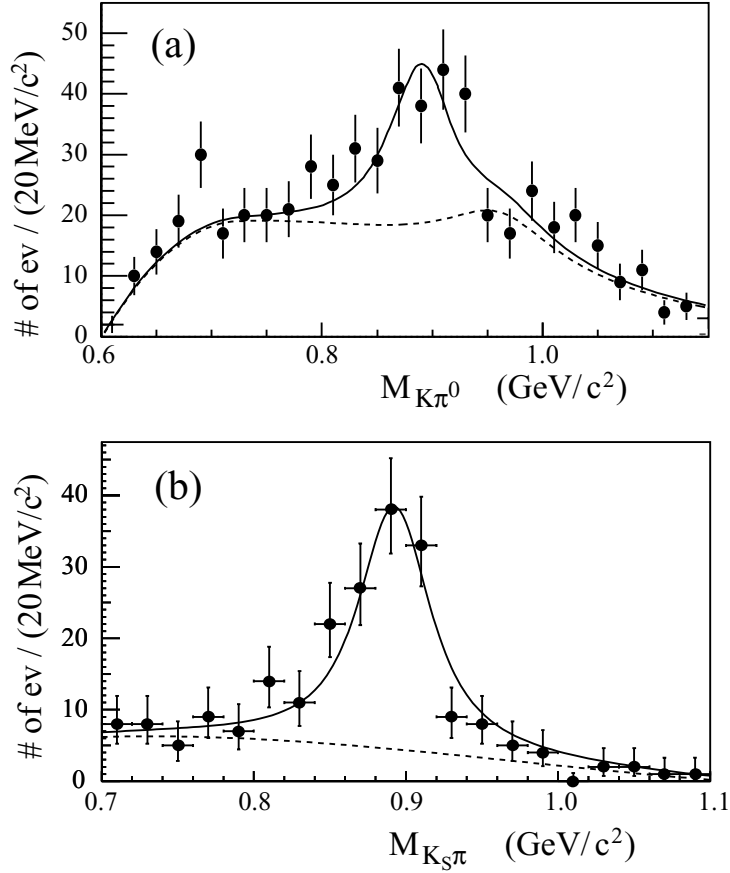}}}
\caption{
(a) $K^-\pi^0$ invariant mass distribution for $K^-\pi^0\eta\nu_{\tau}$
events and (b) the $\pi^- K_S^0$ invariant mass distribution for 
$\pi^- K_S^0 \eta \nu_\tau$ events. Data are fitted with a convoluted 
Breit-Wigner function, and the best 
fit is indicated by the solid 
curve, while 
the BG component is shown by the dashed curve. 
The fits give $N_{K^{*-}\eta\nu_{\tau}}=122\pm 17$ events (a) and 
$N_{K^{*-}\eta\nu_{\tau}}=123\pm 14$ (b), respectively.
}
\label{fig:fit_final.eps}
\end{figure}

The $(K \pi)^-$ invariant mass spectra in both $\tau^- \to K^-\pi^0\eta\nu_
{\tau}$ and $\tau^- \to \pi^- K_S^0 \eta\nu_{\tau}$ decays are analyzed 
to determine the $K^*(892)^-$ content in the final state.

\subsubsection{$\tau^- \to K^-\pi^0\eta\nu_{\tau}$ samples}\label{K*2K-pi0}

To select $\tau^- \to K^-\pi^0\eta\nu_{\tau}$, the $\eta$ signal is
identified with the criterion
$0.50~{\rm GeV}/c^2<M_{\gamma\gamma}<0.58~{\rm GeV}/c^2$,
and then the $K^- \pi^0$
invariant mass $M_{K^-\pi^0}$ distribution is analyzed
(see Fig.~\ref{fig:fit_final.eps} (a)). 
The contribution of the combinatorial
BG under the $\eta$ peak is evaluated from 
the sidebands with
$0.43~{\rm GeV}/c^2<M_{\gamma\gamma}<0.48~{\rm GeV}/c^2$ and 
$0.60~{\rm GeV}/c^2<M_{\gamma\gamma}<0.65~{\rm GeV}/c^2$.
The dashed curve in Fig.~\ref{fig:fit_final.eps} (a)
indicates the BG component estimated from the above samples and from MC
that includes a $\tau$ decay with an $\eta$.
A MC study indicates that the enhancement in the
higher $M_{K^-\pi^0}$ mass region mostly
arises from $\pi^-\pi^0\eta\nu_{\tau}$ events through misidentification
of $\pi^-\to K^-$, while 
the dominant BG, mostly populating the lower mass region, is composed of
contributions from generic $\tau$ decays. 
A clear $K^*(892)^-$ signal over BG can be seen in 
the spectrum shown in Fig.~\ref{fig:fit_final.eps} (a). 

The observed $M_{K^-\pi^0}$ spectrum is fitted with a $K^*$ Breit-Wigner (BW)
function plus a BG component, where the former is convoluted with 
a response function with a mass resolution of 
$\sigma_{M_{K^-\pi^0}}$ = 13.5 MeV/c$^2$.  
The latter is fixed so as to reproduce the $M_{K^-\pi^0}$ spectrum of 
the sidebands.  
The mass and width of the BW function are fixed to those of the
$K^*(892)^-$~\cite{PDG} in the fit. 
The best-fit result is indicated by the solid curve in the figure, and 
gives $N_{K^{*-}\eta\nu_{\tau}}=122 \pm 17$ events with 
a $\chi^2/{\rm d.o.f} = 27.4 / 27$. 

We examine the non-resonant $K^-\pi^0$ contamination by 
adding an additional term in the fit, assuming a phase-space
distribution with a $V-A$ weak interaction
for a hadronic final system.
The fit gives $N_{K^{*-}\eta\nu_{\tau}}=104 \pm 20$ events and 
$N_{\rm {non-}K^{*-}}=42 \pm 25$ events with a $\chi^2/{\rm d.o.f} = 24.5 / 26$.
Since no significant difference between the two fits is found within 
the errors, we here simply take the intermediate state to be purely 
$K^{*-}\eta\nu_{\tau}$ and the difference is taken into account as
the systematic uncertainty for the fitting function.
Even allowing for interference between the $K^{*-}$ and non-resonant $K^-\pi^0$,
the difference is found to be negligible.

No significant $K^*(892)^-$ BG contribution is found in 
generic $\tau$ decays, while $6.5\pm 2.3$ events are estimated 
from the $q\bar{q}$ MC. 

The detection efficiency is evaluated from MC as $\epsilon=0.12\%$, 
including  the branching fractions for $K^*(892)^-\to K^-\pi^0$ and 
$\eta\to\gamma\gamma$ decays. 

The systematic uncertainties in the evaluation of 
${\cal B}(K^*(892)^-\eta\nu_{\tau})$ 
are summarized in Table \ref{Table:sysKstar}. 
The dominant error arises from the uncertainty in the 
fitting function discussed above. It amounts to 15\%.  
Other sources of systematic uncertainties are the same as those for
$K^-\pi^0\eta\nu_{\tau}$ decay. 
The total systematic uncertainty is 16.2\%. 

Thus we obtain the branching fraction 
\begin{eqnarray}
{\cal B}(\tau^-\to K^*(892)^-\eta\nu_{\tau})_{K^{*-}\to K^- \pi^0}
\hspace*{1.5 cm} \nonumber\\
= (1.13\pm 0.17\pm 0.18) 
\times 10^{-4}.  
\end{eqnarray}

For  the non-resonant $K^-\pi^0 \eta\nu_{\tau}$ decay, 
we set an upper limit on its branching fraction  
\begin{eqnarray}
{\cal B}(\tau^-\to K^-\pi^0\eta\nu_{\tau})_{\rm non-resonant} < 3.5\times 10^{-5}
\end{eqnarray}
at the 90\% CL, assuming a pure phase-space distribution 
for the final hadronic system.

\begin{table}[t]
 \caption{Summary of the systematic uncertainties in
 $K^{*-}\eta\nu_{\tau}$ analysis (\%).}
 \label{Table:sysKstar}
 \begin{center}
 \begin{tabular}{c|cc}\hline
 $K^*(892)^-$ decay mode & $K^-\pi^0$ & $K_S^0 \pi^-$ \\
\hline \hline 

BG subtraction &  & \\
$q\bar{q}$     & 2.0 & $-$ \\
\hline
%
Detection efficiency & &  \\
$K/\pi$- /lepton-id              &  2.2/ 2.5 & $-$/ 2.4 \\
Tracking                         &  1.3      & 3.3 \\
$\pi^0/ \eta \to \gamma\gamma$   &  2.0/ 2.0 & $-$/ 2.0 \\
$\pi^0$-veto                     &  2.8      & $-$ \\
\hline
%
Stat. error of signal MC        & 1.7 & 0.7 \\
${\cal B}(K_S^0\to \pi^+\pi^-)$   & $-$ & 0.1 \\
Mass spectrum                   & 0.5 & 0.5 \\
\hline
Luminosity meas. & \multicolumn{2}{c}{1.4}  \\
$\sigma(e^+e^-\to\tau^+\tau^-)$  & \multicolumn{2}{c}{0.3} \\
\hline
%
Fitting function&              15.0 & 4.1 \\
\hline \hline
Total                            &  16.2  &  6.4  \\
    \hline
  \end{tabular}
 \end{center}
\end{table}


\subsubsection{$\tau^-\to K_S^0\pi^-\eta\nu_{\tau}$ samples}\label{K*2pi-Ks}

The $M_{K_S^0 \pi^-}$ distribution 
is shown in Fig.~\ref{fig:fit_final.eps}(b) and is similar to the
$M_{K^-\pi^0}$ case.
A clear $K^*(892)^-$ signal can be seen over a small continuum BG. 

Assuming no non-resonant $K_S^0\pi^-$ contribution, 
the distribution is fitted with a convoluted
BW function plus a BG, composed of a 
third-order polynomial, applying the same method used
for the $K^-\pi^0\eta\nu_{\tau}$ sample.
The yield of $K^*(892)^-$ is $123\pm 14$ events. 
The $K^{*-}$ BG is estimated to be 
$14.5 \pm 5.0$ events from the $q\bar{q}$ continuum, and is subtracted 
from the above yield.

We examine a possible non-resonant $K_S^0\pi^-$ contribution, whose 
mass spectrum is calculated by MC, assuming a pure hadronic phase-space
distribution with a $V-A$ weak interaction. 
Including the non-resonant component, the best fit gives
$N_{K^{*-}} = 121\pm 16$ events and $N_{\rm non-resonant} = 3\pm 15$ events. 
Therefore, we give results assuming no non-resonant background.

The detection efficiency is evaluated by MC as $\epsilon = 0.10\%$, 
which includes the relevant branching fractions:
${\cal B}(K^*(892)^-\to \pi^- K_S^0)$, ${\cal B}(K_S^0\to\pi^+\pi^-)$,
and ${\cal B}(\eta\to\gamma\gamma)$.

The systematic uncertainties are summarized in Table \ref{Table:sysKstar}.
Their magnitudes are similar to those in the $K^-\pi^0\eta\nu_{\tau}$ case, 
except for the non-resonant contribution.  
The total systematic uncertainty is 6.4$\%$.
Consequently, the branching fraction is
\begin{eqnarray}
{\cal B}(\tau^-\to K^*(892)^-\eta\nu_{\tau})_{K^{*-}\to \pi^- K_S^0}
 \hspace*{1.5 cm}\nonumber \\
= (1.46\pm 0.16\pm 0.09) \times 10^{-4}.
\end{eqnarray}

Two measurements using $K^*(892)^-\to K^-\pi^0$ and 
$K_S^0\pi^-$ decays are in agreement, in accordance with isospin symmetry.  
Therefore we combine the two results and obtain
\begin{eqnarray}
{\cal B}(\tau^-\to K^*(892)^-\eta\nu_{\tau}) \hspace*{2.5 cm} \nonumber \\
= (1.34 \pm 0.12 \pm 0.09) \times 10^{-4}. 
\end{eqnarray}


\section{Result}\label{conclusion}

\begin{table}[t]
 \caption {Comparison with previous results. }
 \label{Table:Comparison}
 \begin{center}
  \begin{tabular}{c|c|c|c}
     \hline
 & \multicolumn{2}{c}{Branching fraction ${\cal B}$ $(\times 10^{-4})$} & \\
\hline\hline 
  Mode & This work & Previous exp.  & Reference \\
     \hline
$\tau^- \to K^- \eta \nu_{\tau}$  & $1.58\pm0.05\pm0.09$ & $2.6\pm 0.5\pm 0.5$ &CLEO~\cite{CLEOketanu}\\
 & & $2.9\pm 1.3 \pm 0.7$ &ALEPH~\cite{ALEPH}\\
     \hline
$\tau^- \to \pi^-\pi^0 \eta \nu_{\tau}$  & $13.5\pm 0.3\pm 0.7$ & 
$17\pm 2\pm 2$&CLEO~\cite{CLEOpipi0etanu} \\
                        & & $18\pm4\pm 2$&ALEPH~\cite{ALEPH} \\
     \hline
$\tau^- \to K^-\pi^0 \eta \nu_{\tau}$  & $0.46\pm0.11\pm0.04$ & 
$1.77\pm0.56\pm0.71$&CLEO~\cite{CLEOkpi0etanu}\\
     \hline
$\tau^- \to \pi^- K_S^0 \eta \nu_{\tau}$  & $0.44\pm 0.07\pm 0.02$ & 
$1.10\pm0.35\pm0.11$ &CLEO~\cite{CLEOkpi0etanu} \\
     \hline
$\tau^- \to K^{*-} \eta \nu_{\tau}$  & $1.34\pm 0.12\pm 0.09$ &
$2.90\pm 0.80\pm 0.42$&CLEO~\cite{CLEOkpi0etanu} \\
      \hline
  \end{tabular}
 \end{center}
\end{table}

Using a high statistics 450 million $\tau$-pair data sample from Belle, we
have obtained the following branching fractions 
for five different decay modes:
\begin{eqnarray}
&&{\cal B}(\tau^-\to K^-\eta\nu_{\tau}) = (1.58\pm 0.05\pm0.09)\times 10^{-4}, \nonumber\\
&&{\cal B}(\tau^-\to \pi^-\pi^0\eta\nu_{\tau}) = (1.35\pm 0.03\pm 0.07)\times 10^{-3}, \nonumber\\
&&{\cal B}(\tau^-\to K^-\pi^0\eta\nu_{\tau}) = (4.6\pm 1.1\pm 0.4)\times 10^{-5},\nonumber\\
&&{\cal B}(\tau^-\to \pi^- K_S^0 \eta\nu_{\tau}) = (4.4\pm 0.7\pm 0.3)\times 10^{-5}, \nonumber\\
&&{\cal B}(\tau^-\to K^*(892)^-\eta\nu_{\tau}) = (1.34 \pm 0.12 \pm 0.09)\times 10^{-4}, \nonumber 
\end{eqnarray}
where the first and second errors are  statistical and systematic, 
respectively. 
We also set the upper limits on the following decay modes at the 90\% CL: 
\begin{eqnarray}
&&{\cal B}(\tau^-\to K^- K_S^0 \eta\nu_{\tau}) < 4.5\times 10^{-6},\nonumber\\
&&{\cal B}(\tau^-\to \pi^- K_S^0 \pi^0\eta\nu_{\tau}) < 2.5\times 10^{-5}, \nonumber\\
&&{\cal B}(\tau^-\to K^- \eta\eta\nu_{\tau}) < 3.0\times 10^{-6}, \nonumber\\
&&{\cal B}(\tau^-\to \pi^-\eta\eta\nu_{\tau}) < 7.4 \times 10^{-6}, \nonumber \\
&&{\cal B}(\tau^- \to K^-\pi^0\eta\nu_{\tau})_{\rm non-resonant} < 3.5\times 10^{-5}. \nonumber
\end{eqnarray}

In Table \ref{Table:Comparison}, our results are compared to those previously
obtained by the CLEO~\cite{CLEOpipi0etanu,CLEOketanu,CLEOkpi0etanu} and
ALEPH~\cite{ALEPH} collaborations. 
It is clearly seen that the precision of the measured values
has been considerably improved.

It is also noteworthy that the central values of our branching fractions 
are in all modes lower than those of the other 
experiments~\cite{CLEOpipi0etanu,CLEOketanu,CLEOkpi0etanu,ALEPH}. 
This fact can be mostly attributed to the underestimation
of the BG contamination 
in previous low statistics measurements. 
For instance, 
the $q\bar{q}$ BG estimation in the previous analyses relied on the MC.
However, this analysis evaluates the BG rate using 
a $q\bar{q}$ enriched data sample.
Furthermore,
Ref.~\cite{CLEOketanu} ignored the $K^-\pi^0\eta\nu_{\tau}$ BG 
contribution, since they estimated it as $\sim 0.1\%$ BG, based on a theoretical
calculation that predicted its branching fraction to be 
$8.8\times 10^{-6}$ \cite{Pich}. 
However, its rate is about 1\% of the signal in our measurement, and 
its branching fraction is
about five times as large.  
Our high statistics allows us to reliably and precisely 
scrutinize various BG contributions, while the previous measurements,
which had low statistics, did not have sufficient sensitivity to notice, or 
correctly judge, the BG contamination.

\vspace*{2 mm}

Our results are compared with different theoretical calculations 
in Table \ref{Table:tab}.  
The branching fraction for $\tau^-\to\pi^-\pi^0\eta\nu_{\tau}$ decay 
can be predicted by the isospin symmetry (CVC) using the experimental 
results on 
$e^+e^-\to\pi^+\pi^-\eta$~\cite{CVC,Gilman}. 
Our branching fraction agrees with the predictions, 
${\cal B}(\tau^-\to \pi^-\pi^0\eta\nu_{\tau}) = (1.3\pm 0.2)\times 10^{-3}$ 
\cite{CVC} and $1.5 \times 10^{-3}$ \cite{Gilman}.

Pich~\cite{Pich}, Braaten~\cite{Braaten}, Li~\cite{Li} and 
Aubrecht~\cite{Aubrecht} 
calculated the branching fractions of various $\tau$ decays 
involving $\eta$ meson(s), 
based on chiral perturbation theory, as listed in 
Table~\ref{Table:tab}. 
Their results are in fair agreement with our measurements,
particularly taking into 
account possible uncertainties in theoretical predictions.
Further detailed studies of the physical dynamics in $\tau$ decays
with $\eta$ mesons are required. 

\vspace*{2 mm}

\begin{figure*}[tb]
\centerline{
\resizebox{0.9\textwidth}{!}{%
\includegraphics{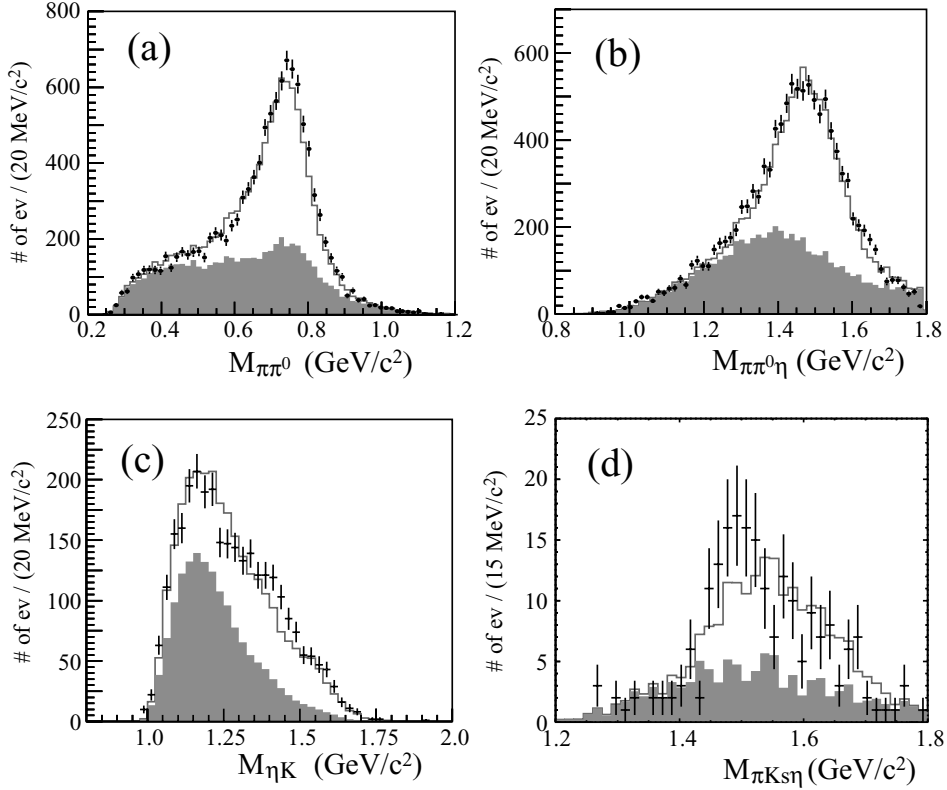}}}
\caption{Invariant mass distributions of (a) $\pi^- \pi^0$
and (b) $\pi^- \pi^0 \eta$ for $\tau^-\to\pi^-\pi^0\eta\nu_{\tau}$ decay,
(c) $K^- \eta$ for $\tau^-\to K^-\eta\nu_{\tau}$ and
(d) $\pi^- K_S^0 \eta$ for $\tau^-\to K^{*-}\eta\nu_{\tau}$,
 $K^{*-}\to K_S^0 \pi^-$.
The points with error bars are the data. 
The normal and filled histograms indicate the signal and $\tau\tau$ BG
MC distributions, respectively. 
The $q\overline{q}$ BG is strongly suppressed and negligible in our sample.
}
\label{fig:mass_pipi0eta.eps}
\end{figure*}

It should be mentioned that the TAUOLA MC 
program~\cite{KKMC} qualitatively reproduces our 
hadronic mass distributions in $\tau^-\to \pi^-\pi^0\eta\nu$, 
$\tau^-\to K^-\eta\nu$ and $\tau^-\to K^*(892)^-\eta\nu$ decays, 
as shown in Fig.~\ref{fig:mass_pipi0eta.eps}.
Further analysis of the spectra including various phenomenological models
is in progress.

\begin{table*}[tb]
\rotatebox{90}{\begin{minipage}{\textheight}
\centering
 \caption {Comparison with theoretical calculations}
 \label{Table:tab}
  \begin{tabular}{c|c|ccccc}
     \hline
      & \multicolumn{6}{c}{Branching fractions} \\
\hline
 Mode & This work & Pich~\cite{Pich} & Gilman~\cite{Gilman} & Braaten~\cite{Braaten} & 
 Li~\cite{Li} & Aubrecht~\cite{Aubrecht} \\
\hline\hline 
\multicolumn{1}{l}{$\tau^- \to K^- \eta \nu_{\tau}$}  &  
  \multicolumn{1}{|l|}{$(1.58\pm 0.05\pm 0.09)\times 10^{-4}$} &
   $1.2\times 10^{-4}$ &  &  & $2.2\times 10^{-4}$ & $1.6\times 10^{-4}$ \\
\multicolumn{1}{l}{$\tau^- \to \pi^-\pi^0 \eta \nu_{\tau}$}  & 
  \multicolumn{1}{|l|}{$(1.35\pm 0.03\pm 0.07)\times 10^{-3}$} &
    $3\times 10^{-3}$ & $1.5\times 10^{-3}$ & $1.4\times 10^{-3}$ & $1.9\times 10^{-3}$ &    \\
\multicolumn{1}{l}{$\tau^- \to K^-\pi^0 \eta \nu_{\tau}$}  & 
  \multicolumn{1}{|l|}{$(4.6\pm 1.1\pm0.4)\times 10^{-5}$} &
   $8.8\times 10^{-6}$ &   &   &   &  $7.6\times 10^{-6}$   \\
\multicolumn{1}{l}{$\tau^- \to \pi^- K_S^0  \eta \nu_{\tau}$}  & 
  \multicolumn{1}{|l|}{$(4.4\pm 0.7\pm 0.3)\times 10^{-5}$} &
   $1.1\times 10^{-5}$ &   &   &   &  $1.0\times 10^{-5}$   \\
\multicolumn{1}{l}{$\tau^- \to K^{*-} \eta \nu_{\tau}$}  & 
  \multicolumn{1}{|l|}{$(1.34\pm 0.12\pm 0.09)\times 10^{-4}$} &
     &   &   & $1.0\times 10^{-4}$ &    \\ 
\multicolumn{1}{l}{$\tau^- \to K^- K_S^0 \eta \nu_{\tau}$}  & 
  \multicolumn{1}{|l|}{$< 4.5 \times 10^{-6}$} &
    $1.6\times 10^{-7}$ &   &   &   & $1.4\times 10^{-7}$ \\
\multicolumn{1}{l}{$\tau^- \to \pi^- K_S^0 \pi^0 \eta \nu_{\tau}$}  & 
  \multicolumn{1}{|l|}{$< 2.5 \times 10^{-5}$} &
       &  &  &  &  \\
\multicolumn{1}{l}{$\tau^- \to K^- \eta\eta \nu_{\tau}$}  & 
  \multicolumn{1}{|l|}{$< 3.0 \times 10^{-6}$} &
    $1.6\times 10^{-9}$ &   &   &   &  $6.9\times 10^{-9}$ \\
\multicolumn{1}{l}{$\tau^- \to \pi^- \eta\eta \nu_{\tau}$}  & 
  \multicolumn{1}{|l|}{$< 7.4 \times 10^{-6}$} &
    $1.1\times 10^{-9}$ &  &  &  &  \\
\hline
\end{tabular}
\end{minipage}
}
\end{table*}


\bigskip

{\bf Acknowledgments}
\smallskip

We thank the KEKB group for the excellent operation of the
accelerator, the KEK cryogenics group for the efficient
operation of the solenoid, and the KEK computer group and
the National Institute of Informatics for valuable computing
and SINET3 network support as well as Tau lepton physics research 
center of Nagoya University. We acknowledge support from
the Ministry of Education, Culture, Sports, Science, and
Technology of Japan and the Japan Society for the Promotion
of Science; the Australian Research Council and the
Australian Department of Education, Science and Training;
the National Natural Science Foundation of China under
contract No.~10575109 and 10775142; the Department of
Science and Technology of India; 
the BK21 program of the Ministry of Education of Korea, 
the CHEP src program and Basic Research program (grant 
No. R01-2008-000-10477-0) of the 
Korea Science and Engineering Foundation;
the Polish State Committee for Scientific Research; 
the Ministry of Education and Science of the Russian
Federation and the Russian Federal Agency for Atomic Energy;
the Slovenian Research Agency;  the Swiss
National Science Foundation; the National Science Council
and the Ministry of Education of Taiwan; and the U.S.\
Department of Energy. This work is supported by
a Grant-in-Aid for Science Research on Priority Area (New 
Development of Flavor Physics) from the Ministry of Education, 
Culture, Sports, Science and Technology of Japan and Creative 
Scientific Research (Evolution of Tau-lepton Physics) from the 
Japan Society for the Promotion of Science.


\begin{thebibliography}{99}
\bibitem{UL} 
   The upper limit is estimated by following the method described in 
   Ref.~\cite{PDG} section 32.3.2.4.
\end{thebibliography}
\end{document}